\documentclass[12pt,aps,prl,superscriptaddress,showpacs,preprintnumbers,amsmath,amssymb, footinbib, longbibliography]{revtex4-2}

\newcommand{\RN}[1]{%
  \textup{\uppercase\expandafter{\romannumeral#1}}%
}

\usepackage{lineno}
\usepackage{graphicx} 
\usepackage{dcolumn}  
\usepackage{epstopdf}
\usepackage{color}
\usepackage[bookmarks]{hyperref}
\usepackage{multirow}

\graphicspath{{ps}}

\newcommand{\ra}{\rightarrow}
\newcommand{\lambdac}{\Lambda^{+}_{c}}

\newcommand{\Kminus}{K^{-}}

\newcommand{\piplus}{\pi^{+}}
\newcommand{\pkpi}{\lambdac \ra p \Kminus \piplus}

\newcommand{\fb}{\rm fb^{-1}}
\newcommand{\pk}{pK^{-}}
\newcommand{\el}{\Lambda\eta}

\newcommand{\emev}{\rm MeV}

\newcommand{\mmev}{{\rm MeV}/c^{2}}

\newcommand{\mgev}{{\rm GeV}/c^{2}}

\newcommand{\flatte}{Flatt\'{e}~}

\begin{document}


\title{ \quad\\[1.0cm] Observation of a Threshold Cusp at the $\el$ Threshold in the $\pk$ Mass Spectrum with $\pkpi$ Decays}


\collaboration{The Belle Collaboration}

\begin{abstract}
 We observe a narrow peaking structure in the $\pk$
invariant-mass spectrum near the $\el$ threshold. 
The peak is clearly seen in 1.5 million events of 
$\pkpi$ decay using the $980~\fb$ data sample collected 
by the Belle detector at the KEKB asymmetric-energy 
$e^{+}e^{-}$ collider. 
We try two approaches to explain this structure: 
as a new resonance
and as a cusp at the $\el$ threshold. 
The best fit is obtained with a coherent sum of  
a Flatt$\rm{\acute{e}}$ function and a constant background amplitude
with the $\chi^2/{\rm n.d.f}=257/243$ ($p = 0.25$),
while the fits to Breit-Wigner functions are unfavored by more than 7$\sigma$.
The best fit explains the structure as a cusp at the $\el$ threshold
and the obtained parameters are consistent with the known
properties of $\Lambda(1670)$.
The observation gives the first identification of a threshold cusp
in hadrons from the spectrum shape.
\end{abstract}

\maketitle

\tighten

{\renewcommand{\thefootnote}{\fnsymbol{footnote}}}
\setcounter{footnote}{0}
Regions around the mass thresholds of two hadrons have been 
of great interest for studies of exotic hadrons
such as $X(3872)$ and $P_{c}(4312)^{+}$~\cite{Belle_x3872, LHCb_pc1, LHCb_pc2}, which are
found near mass thresholds of two hadrons.
These near-threshold resonances could appear as threshold cusps 
instead of usual smooth peaks with Breit-Wigner (BW) shape.
A cusp, defined as a discontinuity in the derivative
of spectrum function, always appears exactly at the threshold,
and its position does not reflect the pole position of a resonance~\cite{cusp_theory}.
To understand the nature of a near threshold behavior, 
it is necessary to identify whether the peak structure
is a threshold cusp or usual peak of BW type.
In principle, a threshold cusp can be distinguished from a smooth
peak because the derivative diverges at the peak position, but
practically, experimental mass resolution often makes
such identification difficult~\cite{LHCb_line_shape_x3872}.
Therefore, there are just a few cases where threshold cusp is
identified~\cite{cusp1_na48, cusp2_na48, cusp3_a2, cusp4_alice, cusp5_cbelsa},
and none of them are from the spectrum shape.

In this Letter, we report a newly discovered peaking structure 
in the $\pk$ mass spectrum near the $\el$ mass threshold
\footnote{Unless stated otherwise, charge-conjugate modes are implied throughout this Letter.}.
A trace of this peak structure is observed in the previous analysis~\cite{DCSD_Belle} of 
$\pkpi$ decay using a 980 $\fb$ data sample collected by the Belle Collaboration.
A similar structure is also seen by LHCb
in the same $\Lambda_{c}^{+}$ decay channel~\cite{LHCb_amplitude}.
We approach this peak considering two possible cases; a BW-type
peak and a visible $\el$ threshold cusp enhanced by the $\Lambda(1670)$ 
pole nearby.

\begin{figure}[!h]
\includegraphics[width=0.5\textwidth]{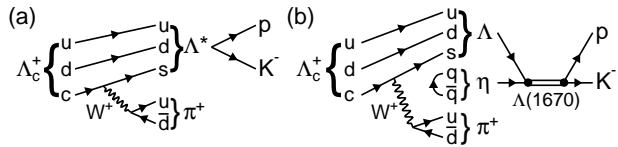}
\caption{Feynman diagrams for (a) a new $\Lambda^{*}$ resonance 
and (b) a visible $\el$ threshold cusp enhanced by the $\Lambda(1670)$ pole 
in $\pkpi$ decay. 
}
\label{fig:feynman}
\end{figure}

If it is a BW-type peak, it suggests an existence of
a new resonance. In this regard,
two theory groups independently proposed a narrow
$\Lambda^{*}$ resonance with spin 3/2 near the $\el$ threshold \cite{Liu_pwa, Kamano_pwa2}
based on the $\pk\ra\el$ data \cite{CrystalBall}, and
the peak could be due to this $\Lambda^{*}$ resonance in the $\pkpi$ decay 
as shown in Fig.~\ref{fig:feynman}(a).
Such an exotic state is not expected in the quark model, 
and thus it is important to study the observed peak structure
to see whether it is the case or not.

On the other hand, a visible cusp can arise 
via the $\el$ rescattering process in the $\pkpi$ decay
as shown in Fig.~\ref{fig:feynman}(b).
In this case, the $\Lambda(1670)$ could be involved in
the $S$-wave $\eta\Lambda$-$pK^{-}$ rescattering.
Therefore, the shape of the peaking structure is determined
by the properties of $\Lambda(1670)$ 
such as partial widths of the $\Lambda(1670)$ into $pK^{-}$ and $\eta\Lambda$ channels.

In this analysis, we use data collected by the Belle detector 
at the KEKB asymmetric-energy $e^{+}e^{-}$ collider~\cite{KEKB1, *KEKB2}.
The data sample is taken at or near the $\Upsilon(nS)$ ($n=$1-5) resonances.
The Belle detector is a large-solid-angle magnetic spectrometer 
consisting of a silicon vertex detector (SVD)~\cite{Belle_SVD1, *Belle_SVD2}, a central drift chamber (CDC), 
an array of aerogel threshold Cherenkov counters (ACC), 
barrel-like arrangement of time-of-flight scintillation counters (TOF), 
and an electromagnetic calorimeter composed of CsI(Tl) crystals (ECL) 
located inside a superconducting solenoid coil with a 1.5 $\rm T$ magnetic field. 
The detector is described in detail elsewhere \cite{Belle_detector1, *[{also see detector section in }] Belle_detector2}.
\par
We also use samples of $e^{+}e^{-} \ra c\bar{c}$ Monte Carlo (MC) events 
to estimate reconstruction efficiencies and detector performance. 
The MC simulation samples are generated with PYTHIA~\cite{PYTHIA} and EVTGEN~\cite{EVTGEN1, *EVTGEN2} and propagated by GEANT3~\cite{GEANT3}.

The same event selection criteria as in the previous $\pkpi$ analysis~\cite{DCSD_Belle} 
are used to reconstruct the decay event from the charged $p$, $K^{-}$, and $\pi^{+}$. 
All charged tracks have a distance-of-closest-approach to the
interaction point of less than 2.0 cm in the beam direction ($z$) 
and less than 0.2 cm in the transverse ($r$-$\phi$) direction. 
They are also required at least one SVD hit.
The particle identification (PID) likelihoods $\mathcal{L}(h)$ 
($h$ is $p^{\pm}$, $K^{\pm}$, or $\pi^{\pm}$) are derived from 
measurements using the CDC, TOF, and ACC~\cite{Belle_PID}. 
The ratio of likelihoods, $\mathcal{R}(h:h^{'})$, is defined as 
$\mathcal{L}(h)/[\mathcal{L}(h) + \mathcal{L}(h^{'})]$ for $h$ and $h^{'}$ identification.
The PID requirements for the three charged hadrons are 
$\mathcal{R}(p:K) > 0.9$ and $\mathcal{R}(p:\pi) > 0.9$ for $p$, 
$\mathcal{R}(K:p) > 0.4$ and $\mathcal{R}(K:\pi) > 0.9$ for $K^{-}$, 
and $\mathcal{R}(\pi:p) > 0.4$ and $\mathcal{R}(\pi:K) > 0.4$ for $\pi^{+}$.
In addition, the electron likelihood ratio derived from ACC, CDC, 
and ECL measurements is required to be less than 0.9 for all hadrons~\cite{Belle_EPID}.
To reduce combinatorial backgrounds, we require a scaled momentum,
defined as $p^{*}/\sqrt{E_{\rm cm}^{2}/4-M^{2}}$, to be greater than 0.53; 
here, $p^{*}$, $E_{\rm cm}$, and $M$ are the $\Lambda_{c}^{+}$ momentum in the center of mass frame, 
the total center-of-mass energy, and the mass of the $\Lambda_{c}^{+}$ candidate, respectively.
The three charged tracks are fitted to a common vertex, 
and the $\chi^{2}$ value of the vertex fit is required to be less than 40.
The $1.5 \times 10^{6}$ $\pkpi$ decays are reconstructed with
these event selection criteria.
For removing non-$\Lambda_{c}^{+}$ backgrounds, we subtract events 
in the signal range, $2.2746 < M(pK^{-}\pi^{+}) < 2.2986~\mgev$, 
by events in the sideband ranges, $2.2506<M(pK^{-}\pi^{+})<2.2626~\mgev$ 
and $2.3106<M(pK^{-}\pi^{+})<2.3226~\mgev$.
\par
To improve the invariant-mass resolution on the $M(pK^{-})$ distribution, 
three daughter particles of the decay are fitted to the common vertex point 
with the mass of $\Lambda_{c}^{+}$.
After this mass-constraint vertex fit, detector responses at 
1663.5 ${\rm MeV}/c^{2}$ on the $M(pK^{-})$ distribution can be 
represented by a double-Gaussian function with a common central mean value.
From a MC simulation, standard deviations of the core and tail Gaussian
functions are determined to be 1.25 $\mmev$ and 2.50 $\mmev$, respectively, 
and the yield of the tail Gaussian function is 0.193 of the core Gaussian function. 
\par

We estimate the reconstruction efficiency of $\pkpi$ decay using the MC sample.
Due to variations of the estimated efficiencies on $M^{2}(K^{-}\pi^{+})$ and $M(pK^{-})$,
we correct the $\Lambda_{c}^{+}$ yields in individual bins of the two-dimensional distribution
of $M^{2}(K^{-}\pi^{+})$ versus $M(pK^{-})$.
\par

\begin{figure}[!h]
\includegraphics[width=0.5\textwidth]{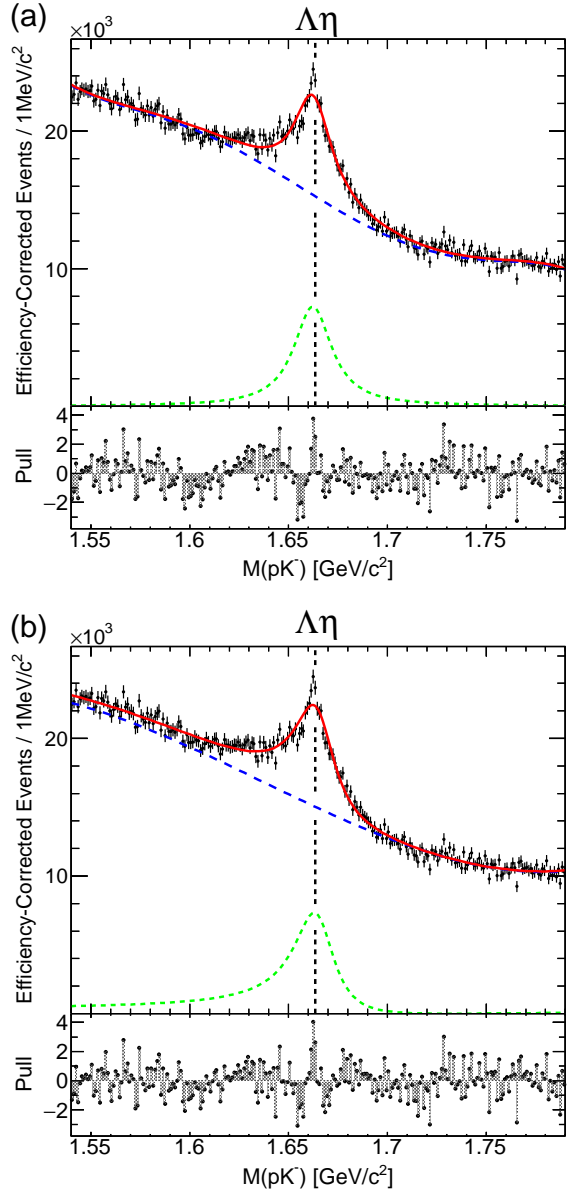}
\caption{
Fits to the $\Lambda_{c}^{+}$ yield in $M(pK^{-})$ spectra with
(a) BW function and (b) BW model to which a complex constant is added.
The curves indicate the full fit model (solid red),
background $\Lambda_{c}^{+}$ decay events (long-dashed blue),
(a) BW peak (dashed green), and (b) BW model with complex constant added coherently (dashed green).
The $\Lambda\eta$ threshold is marked by the vertical dashed lines.
The bottom panels show the pull distributions of the fits.
}
\label{fig:bw}
\end{figure}

From the perspective of the $pK^{-}$ peak as a usual hadron resonance structure, 
we perform a binned least-$\chi^{2}$ fit to the efficiency-corrected $M(pK^{-})$ distribution 
in the range of 1.54 $\mgev$ to 1.79 $\mgev$ with a non-relativistic BW function defined as
\begin{equation}\label{eq:non_BW_eq}
\frac{dN}{dm} \propto 
\left|{\rm BW}(m)\right|^{2} = \left|\frac{1}{(m-m_{0}) 
+ i\frac{\Gamma_{0}}{2} }\right|^{2}, 
\end{equation}
where $m$, $m_{0}$, and $\Gamma_{0}$ are the $pK^-$ invariant mass, the nominal mass, 
and the resonance width, respectively
\footnote{In equation, here and below, we use a natural unit that $c$ equals to 1.}.
The BW function is convolved with the double-Gaussian function 
with fixed parameters to take into account detector responses. 
The probability density function (PDF) for background $\Lambda_{c}^{+}$ decay events
is a fifth-order Chebyshev polynomial function.
Figure~\ref{fig:bw}(a) shows the fit results using the BW function.
The mass and width are obtained to be $1662.4 \pm 0.3~\mmev$ 
and $22.6 \pm 1.5~\emev$, respectively, where the uncertainties are statistical.
The reduced $\chi^{2}$ is 1.35 (328/242).
\par
A better reduced $\chi^{2}$ is obtained by adding
a complex constant to the non-relativistic BW function coherently as 
$\frac{dN}{dm} \propto |{\rm BW}(m) + re^{i\theta}|^{2}$, 
where $r$ and $\theta$ are real parameters,
and $\theta$ is fixed to $\pi$,
leading to constructive interference below the $\Lambda\eta$ threshold
and destructive above that.
Incoherent background $\Lambda_{c}^{+}$ decay events are represented by a third-order Chebyshev polynomial.
Figure~\ref{fig:bw}(b) shows the fit results including the interference
and the mass and width are obtained as $1665.4 \pm 0.5~\mmev$ and $23.8 \pm 1.2~\emev$, respectively,
where the uncertainties are only statistical,
with the reduced $\chi^{2}$ of 1.27 (308/243).

\begin{figure}[!h]
\includegraphics[width=0.5\textwidth]{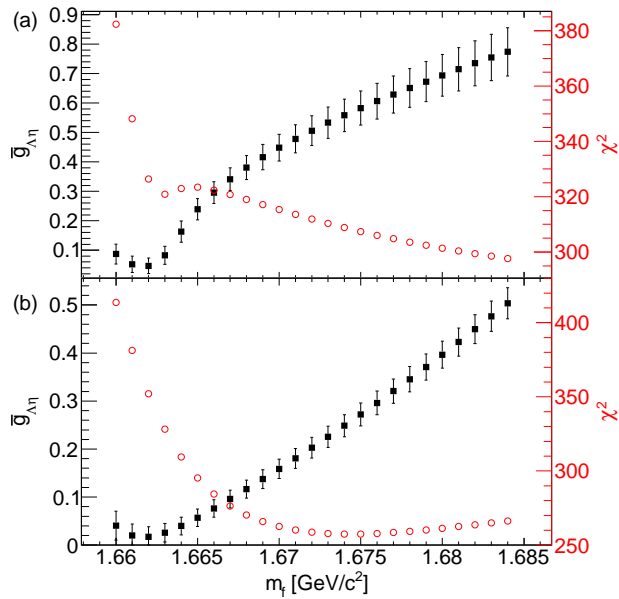}
\caption{
$\bar{g}_{\Lambda\eta}$ and $\chi^{2}$ from \flatte model (a) without and (b) with the interference as a function of fixed $m_{f}$.
The black square and red circle markers indicate $\bar{g}_{\Lambda\eta}$ and $\chi^{2}$, respectively. 
Number of degree of freedom is 242 for all fits in (a) and 243 for all fits in (b). 
Uncertainty of $\bar{g}_{\Lambda\eta}$ is statistical.
}
\label{fig:flatte_mf_all}
\end{figure}

\begin{figure}[!h]
\includegraphics[width=0.5\textwidth]{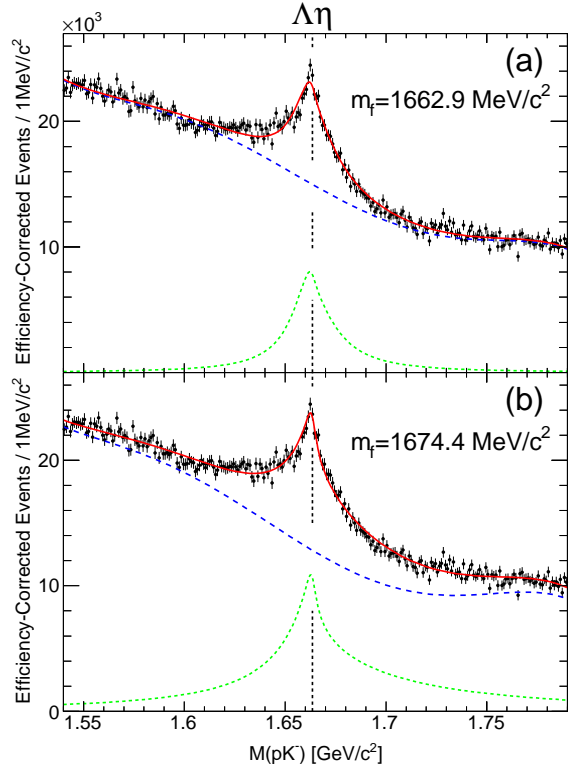}
\caption{
Fits with \flatte function when $m_{f}$ is fixed to (a) $1662.9~{\rm MeV}/c^{2}$
and (b) $1674.4~{\rm MeV}/c^{2}$.
The curves indicate the full fit model (solid red), \flatte function (dashed green), and background $\Lambda_{c}^{+}$ decay events (long-dashed blue).
}
\label{fig:flatte}
\end{figure}

\begin{figure}[!h]
\includegraphics[width=0.5\textwidth]{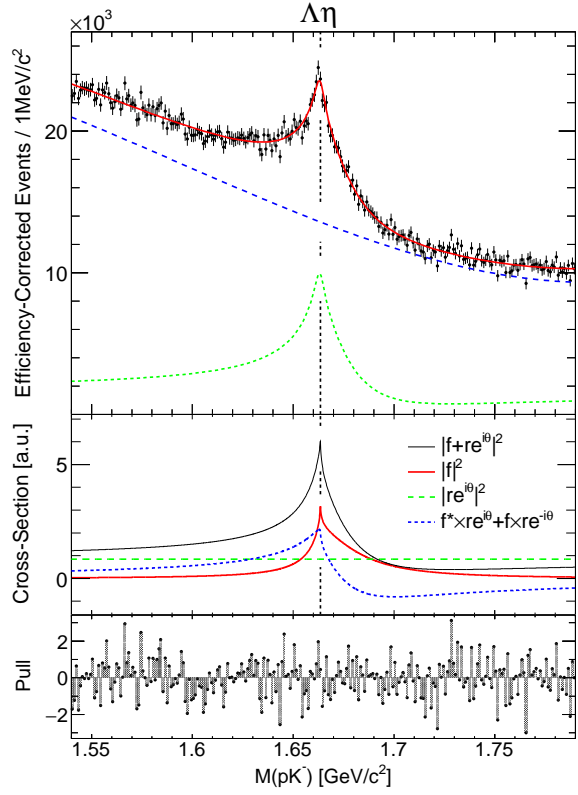}
\caption{
Fit to the $\Lambda_{c}^{+}$ yield in $M(pK^{-})$ spectrum
with \flatte model to which a complex constant is added coherently with $m_{f}=1674.4~{\rm MeV}/c^{2}$ and $\theta=\pi$ being fixed.
In the upper panel, the curves indicate the full fit model (solid red), \flatte function with complex constant added coherently (dashed green), and incoherent background $\Lambda_{c}^{+}$ decay events (long-dashed blue).
The middle panel shows the breakdown for $|f+re^{i\theta}|^{2}$;
the curves indicate the full function (thin solid black), $|f|^{2}$ (thick solid red),
$|re^{i\theta}|^{2}$ (long dashed green), and the interference term (dashed blue). 
The detector response is not taken into account.
The bottom panel shows the pull distribution of the fit.
}
\label{fig:flatte_inters}
\end{figure}

Another possibility is that the peak structure is a cusp at the $\Lambda\eta$ threshold 
enhanced by the $\Lambda(1670)$ pole nearby.
We fit a non-relativistic \flatte function
~\cite{flatte, Baru2005} defined as
\begin{equation} \label{eq:nonrel_flatte}
 \frac{dN}{dm} \propto \left|f(m)\right|^{2} = 
\left|\frac{1}{m-m_{f}+{\frac{i}{2}}\left(\Gamma^{\prime}+\bar{g}_{\Lambda\eta}k\right)}\right|^{2},
\end{equation}
to the peak region, where $m$ is the $pK^-$ invariant mass and $m_{f}$ is 
a parameter corresponding to the nominal mass of $\Lambda(1670)$.
The $\Gamma^{\prime}$ is a parameter for the sum of the partial widths 
of the decay modes other than $\Lambda\eta$,
and is approximated as a constant in the following analysis.
In the formula,
$\bar{g}_{\Lambda\eta}k$ represents the partial decay width of the $\Lambda\eta$ channel,
where $\bar{g}_{\Lambda\eta}$ and $k$ are
the dimensionless coupling constant and the decay momentum in the $\Lambda\eta$ channel,
respectively.
Here, $k$ becomes imaginary below the $\Lambda\eta$ threshold so as to keep the analytic continuity.
We also note that for $\bar{g}_{\Lambda\eta}=0$, Eq. (\ref{eq:nonrel_flatte}) reduces to
the BW function [Eq. (\ref{eq:non_BW_eq})] with $m_f=m_0$ and $\Gamma^{\prime}=\Gamma_{0}$.

Due to the scaling behavior of the \flatte function \cite{Baru2005},
we fix $m_{f}$ when we perform a fit and repeat the fit with various
$m_{f}$ values~\cite{LHCb_line_shape_x3872}.
The signal PDF, \flatte function, is convolved with the double-Gaussian function 
for detector responses and a fifth-order Chebyshev polynomial
represents background $\Lambda_{c}^{+}$ decay events.
Figure~\ref{fig:flatte_mf_all}(a) shows the results on 
$\bar{g}_{\Lambda\eta}$ and $\chi^{2}$ for each fixed $m_{f}$.
Strong correlation between $\bar{g}_{\Lambda\eta}$ and $m_{f}$ is seen 
as expected from the scaling.
Typical fit results with fixed $m_{f}=1662.9~\mmev$
and $1674.4~\mmev$ are shown in Fig.~\ref{fig:flatte}.\par

The best fit is obtained by taking into account an interference
with another $S$-wave amplitude such as a tail of $\Lambda(1405)$.
We take a constant, $re^{i\theta}$, as the amplitude for these $\Lambda_{c}^{+}$ events,
and add it to the \flatte coherently;
here, $\theta$ is simply fixed to $\pi$ to represent the $\Lambda_{c}^{+}$ events
distribution, which drops rapidly above the $\Lambda\eta$ threshold,
and variations from the fixed $\theta$ are considered as a source of systematic uncertainty.

We perform a binned least-$\chi^{2}$ fit with the combined function, 
$\frac{dN}{dm} \propto |f(m) + re^{i\theta}|^{2}$, by changing the fixed $m_{f}$. 
Incoherent background $\Lambda_{c}^{+}$ decay events are represented by a third-order Chebyshev polynomial.
As shown in Fig.~\ref{fig:flatte_mf_all}(b), a strong correlation between 
$m_{f}$ and $\bar{g}_{\Lambda\eta}$ is still seen even when the interference is taken into account.
The best fit with $\chi^{2}/{\rm ndf}=1.06~(257/243)$ is 
obtained at $m_f=1674.4~{\rm MeV}/c^{2}$, and 
the result is shown in Fig.~\ref{fig:flatte_inters}.
$\Gamma^{\prime}$ and $\bar{g}_{\Lambda\eta}$ are determined to
be $27.2 \pm 1.9~\rm MeV$ and $0.258 \pm 0.023$, respectively,
where the uncertainties are only statistical.
The partial width, $\Gamma_{\Lambda\eta}$,
of $\Lambda\eta$ channel is calculated as
the product of $\bar{g}_{\Lambda\eta}$ and $q^{0}_{\Lambda\eta}$,
which is
the center-of-mass momentum of $\Lambda\eta$ at $m = m_{f}$.
Then, the total width, $\Gamma_{\rm tot}$, defined as a sum of $\Gamma^{\prime}$ and $\Gamma_{\Lambda\eta}$ is obtained to be $50.3 \pm 2.9~\emev$, where the uncertainty is only statistical.

\begin{table}[!h]
  \caption
      {
        \label{tbl:flatte_sys_err}
        Systematic uncertainties in $\Gamma^{\prime}$, $\bar{g}_{\Lambda\eta}$, and $\Gamma_{\rm tot}$ from \flatte fit for the $pK^{-}$ peak structure.
      }
      \begin{ruledtabular}
        \begin{tabular}{lccc}
          {Source} & {$\Gamma^{\prime}~{\rm (MeV)}$} & {$\bar{g}_{\el}$ ($\times 10^{-3}$)} & $\Gamma_{\rm tot}~\rm (MeV)$\\
          \colrule
          {Bin size} & $\pm 0.0$ & $\pm 3$ & $\pm 0.3$ \\
          {Detector resolution} & $+0.3,-0.4$ & $+7,-6$ & $\pm 0.2$ \\
          {Absolute mass scale} & $\pm 0.8$ & $+5,-6$ & $\pm 1.3$ \\
          {Fit range} & $+1.1$ & $-36$  & $+0.8, -2.4$ \\
          {Efficiency correction} & $\pm 0.6$ & $\pm 8$ & $\pm 0.2$ \\
          {PDF model} & $+3.5,-1.9$ & $+9,-29$ & $+3.4,-2.1$ \\
          {$\theta$} & $\pm 3.3$ & $\pm 59$ & $\pm 2.0$\\
          \colrule
          {Total} & $+5.0,-3.9$  & $+61,-75$ & $+4.2, -4.0$\\
        \end{tabular}
      \end{ruledtabular}

\end{table}

We estimate the systematic uncertainties for $\bar{g}_{\Lambda\eta}$ and $\Gamma^{\prime}$ 
of the \flatte model with a constant added coherently.
These systematic uncertainties are listed in Table~\ref{tbl:flatte_sys_err}.
We change the bin size of the $M(pK^{-})$ distribution to 2 $\emev$ to check the effect of binning.
Systematic uncertainty from the mass resolution is estimated by 
increasing or decreasing the mass resolution by $20\%$.
The effect of the absolute 
mass scaling is estimated by shifting the overall $M(pK^{-})$ distribution by $\pm 0.2~\mmev$,
which is a difference between a measured $\lambdac$ mass
and the world-average value~\cite{PDG}.
\par
We vary the fit range to estimate
the systematic uncertainty from the choice of the fit range.
The same PDFs are used for fitting to a narrow range from 1.55 $\mgev$ to 1.78 $\mgev$.
In the wide fit range from 1.48 $\mgev$ to 1.8 $\mgev$, 
the peak structure of $\Lambda(1520)$ appears
and is represented by a $D$-wave relativistic BW function convolved with 
a double Gaussian function to represent detector responses.
Background $\Lambda_{c}^{+}$ events are represented by a seventh-order Chebyshev polynomial.
The largest differences in the fit results are considered as the systematic uncertainty from the fit range.
A systematic uncertainty from the efficiency correction is estimated
by performing a fit to the $M(pK^{-})$ distribution without the efficiency correction.
\par
To estimate a systematic uncertainty due to the PDF modelling, we perform the fit with various PDFs.
The PDF for the incoherent $\Lambda_{c}^{+}$ decay events is changed to second and fourth-order Chebyshev polynomials.
We also change the non-relativistic \flatte function to a relativistic form.
In addition, we study a case where all the background $\Lambda_{c}^{+}$ decay events are coherent.
The total PDF is changed to $|f(m)+\sqrt{(p_{0}+p_{1}m+p_{2}m^{2}+p_{3}m^{3})}e^{i\theta}|^{2}$, 
where $p_{i}$s ($i=\rm 0,1,2,~and~3$) and $\theta$ are free parameters.
The largest differences in the fit results of the PDF models are 
taken as the systematic uncertainty from the PDF model. \par

In the \flatte fit, the reduced $\chi^{2}$ is
improved when the interference term is added,
as it reproduces the drop of the background level around the peak structure.
It indicates a significant interference with the background $S$-wave amplitude.
Here we note that resonances in higher partial waves
would not affect the cusp shape, because the discontinuity in the higher partial waves appear only
in the second or higher derivatives, and the interference
with $S$-wave vanishes with an integral over solid angle.
The value of $m_{f}$ that gives the best $\chi^{2}$ is $1674.4~\mmev$,
which is consistent with the recent measurement of $\Lambda(1670)$ mass,
$1674.3 \pm 0.8 \pm 4.9~\mmev$~\cite{Belle_Lam1670}.
The total width at $m_{f}=1674.4~\mmev$ is estimated as $50.3 \pm 2.9 ^{+4.2}_{-4.0}~\emev$
and is also consistent with the recent measurement,
$36.1 \pm 2.4 \pm 4.8 ~\emev$, within $1.9\sigma$ of the total uncertainty. 
In order to determine partial widths of $\Lambda\eta$ and $pK^{-}$
and the \flatte parameters more accurately,
a simultaneous-fit analysis with the $\Lambda(1670)$ peak structure
in the $\Lambda\eta$ distribution is required.
\par
\par
The fit result with the \flatte function to which the constant is
coherently added shows the best reduced $\chi^{2}$
of 1.06 ($257/243$, $p = 0.25$),
in contrast to 1.27 ($308/243$, $p = 3.1 \times 10^{-3}$) from the best BW fit.
In particular, the \flatte function reproduces
the shape near the peak point better than the BW function.
These results show that the present peaking structure
is explained better by a threshold cusp
than to a new hadron resonance by more than 7$\sigma$.
This gives the first identification of a threshold cusp in hadrons from the spectrum shape.
In the cusp interpretation, the structure
near the $\Lambda\eta$ threshold is explained without
the need of a new resonance.
We also note that LHCb explained the structure using
a BW function with fixed mass and width~\cite{LHCb_amplitude}.
A small deviation is observed near the peak structure,
but not considered as significant.
\par
\begin{acknowledgments}
We thank the KEKB group for the excellent operation of the accelerator,
and the KEK cryogenics group for the efficient operation of the solenoid.
\end{acknowledgments}


\bibliography{reference.bbl}

\end{document}